\documentclass[aps,prl,amsmath,amssymb,twocolumn]{revtex4-1}

\usepackage{graphicx}
\usepackage{dcolumn}
\usepackage{bm}
\begin{document}
\title{The Slowdowns of Water Dynamics when Approaching a Glass Transition or a Solid Interface: A Common Rationale}

\author{F. Klameth}
\author{M. Vogel}
\email{michael.vogel@physik.tu-darmstadt.de}
\affiliation{Institut f\"ur Festk\"orperphysik, Technische Universit\"at Darmstadt, Hochschulstr.\ 6, 64289 Darmstadt, Germany}

\date{\today}

\begin{abstract}
Performing molecular dynamics simulations, we investigate the enormous slowdowns of water dynamics when approaching a glass transition or a solid interface. We show that both effects can be described on common grounds within a theoretical framework, which was recently proposed by Schweizer $\emph{et al.}$ and considers coupled local hopping and elastic distortion. For confined water, we correctly describe the variation of the $\alpha$-relaxation time $\tau_\alpha$ as a function of both temperature and position with respect to the interface. Exploiting our knowledge of a cooperative length scale $\xi(T)$ from the confinement studies, we quantitatively rationalize the glassy slowdown, $\tau_\alpha(T)$, and the Stokes-Einstein breakdown of bulk water. For both confined and bulk liquid, variations of the $\alpha$-relaxation time are intimately related to changes of the cage-rattling amplitude.

\end{abstract}


\maketitle

The structural ($\alpha$) relaxation of liquids strongly slows down when a glass transition or a solid interface is approached. In view of an enormous importance in nature and technology, these slowdowns were the subject of numerous research efforts in the last decades \cite{Ediger_96, Angell_00, Debenedetti_01, Cavagna_09}. When cooling a liquid towards a glass transition, characteristic times of the $\alpha$ relaxation increase by almost 15 orders of magnitude in a narrow temperature range so that an Arrhenius law does not apply. It remains a major scientific challenge to explain this tremendous slowdown of liquid dynamics, which is not accompanied by significant changes of the liquid structure. Likewise, when confining a liquid to nanoscopic spaces, its dynamical behavior can be altered substantially \cite{McKenna_05, Vogel_10, Richert_11}. In particular, the time scale of the $\alpha$ relaxation often rises by several orders of magnitude near solid interfaces.

Several theories were put forward to explain the glassy slowdown \cite{Goetze_99, Dyre_06, Wolynes_07, Chandler_10, Berthier_11, Langer_14}. Many workers argued that the enormous temperature dependence, i.e., the high fragility, is due to an increasing length scale. Similarly, despite recent progress \cite{Truskett_08, Franosch_12, Tanaka_11}, a thorough understanding of the effect of confinement on liquids is still lacking. Both research fields are linked by the long tradition to receive information about length scales of glassy dynamics from confined liquids \cite{McKenna_05, Vogel_10, Richert_11}. To avoid spurious liquid-matrix interactions, recent simulation studies pinned appropriate fractions of particles of a bulk liquid so as to obtain confining matrices consisting of the same type of particles as the confined liquid \cite{Tanaka_11, Kob_02, Kob_04, Biroli_08, Biroli_13, Cammarota_12, Cammarota_13, Tarjus_10, Tarjus_12, Kob_12, Berthier_12, Cavagna_12, Hocky_12, Hanson_12, Szamel_13, Vogel_13, Vogel_14}. While the strong slowdown of dynamics near interfaces is accompanied by a substantial perturbation of the structure for most matrices, the slower dynamics occurs in an unperturbed structure for such neutral matrices \cite{Kob_04, Berthier_12, Vogel_13}. In view of these unique properties, neutral confinements not only enable investigations on length scales associated with various theories of the glass transition, but they also provide a key to a fundamental understanding of liquid dynamics at interfaces and in confinements. 

Here, we show that the slowdowns upon cooling or confining a glass-forming liquid can be understood in a common framework that is based on the elastically collective nonlinear Langevin equation (ECNLE) theory \cite{Mirigian_13, Mirigian_14-1, Mirigian_14-2, Mirigian_14-3}. The key quantity of ECNLE theory is a displacement-dependent dynamic free energy, which is related to the effective force exerted on a particle by its surroundings and which is available from the structure factor. ECNLE theory proposes that the $\alpha$ relaxation involves coupled local hopping motion and non-local elastic distortion of the surroundings, resulting in two inter-related, but distinct energy barriers. Specifically, the total free energy barrier $F$ against $\alpha$ relaxation is the sum of a local hopping barrier $F_\text{H}$ and a non-local elastic barrier $F_\text{E}$. While the former prevails at high temperatures, the latter dominates at low temperatures and causes a high fragility. ECNLE theory assumes that the elastic barrier is proportional to a cooperative volume, which grows upon cooling, i.e., a temperature-dependent length scale $\xi(T)$ is involved. It predicts that long-time $\alpha$ relaxation is intimately related to short-time rattling motion within local cages.

Our analyses focus on bulk and confined waters, motivated by their unique relevance in various fields. Specifically, we perform molecular dynamics (MD) simulations for the SPC/E model. We exploit our previous finding that neutral confinements provide access to a relevant length scale of water behavior as they do not affect the hydrogen-bond network \cite{Vogel_13,Vogel_14}. It is shown that use of this length scale in an ECNLE approach allows us to describe not only the position- and temperature-dependent dynamics of confined water, but also the glassy slowdown of bulk water, including a breakdown of the Stokes-Einstein (SE) relation. The present approach focuses on neutral pores of cylindrical shape with 2.5\,nm radius, utilizing the fact that the geometry and size of such confinement hardly affect the length scale. Simulation details can be found in Refs.\ \cite{Vogel_13,Vogel_14}.

First, we investigate a possible relation between long-time and short-time dynamics. The long-time $\alpha$ relaxation is studied based on incoherent scattering functions (ISF) of the oxygen atoms, $S(q,t)$ \cite{Vogel_13,Vogel_14}, where the used modulus of the scattering vector, $q$, corresponds to the oxygen-oxygen next neighbor distance. Characteristic time scales $\tau_\alpha$ are determined from $S(q, t\!=\!\tau_\alpha)=1/\text{e}$. The short-time rattling motion is characterized by mean-square displacements (MSD) of the oxygen atoms, $r_{p}^2\equiv\langle r^2(t=1\,\mathrm{ ps})\rangle$. At $t=1\,$ps and sufficiently low temperatures, the MSD exhibits a plateau, which quantifies the typical amplitudes of rattling motions within local cages formed by neighboring atoms. It yields an estimate of the 'plateau' shear modulus,  $G_p \propto k_\text{B}T/ r_{p}^2$ \cite{Dyre_06, Leporini_12}. 

\begin{figure}
 \centering
 \includegraphics*[width=8.5cm]{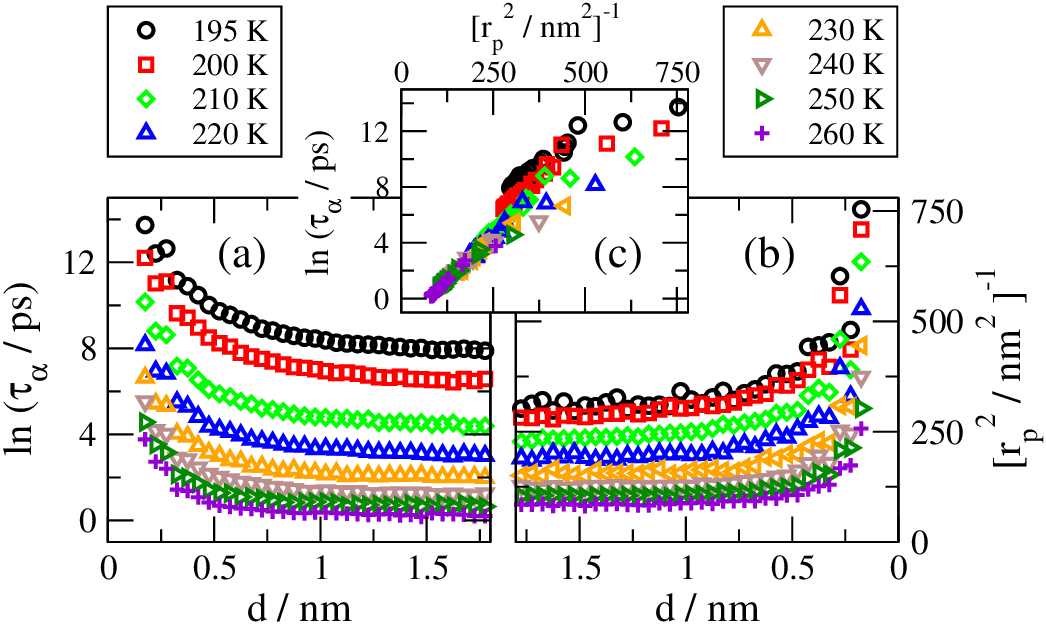}
\caption{Parameters of water dynamics at various temperatures as a function of the distance $d$ to the pore wall: (a) $\ln \tau_\alpha$ and (b) $1/r^2_p$. (c) Dependence of $\ln \tau_\alpha$ on $1/r^2_p$. }
\label{fig1}
\end{figure}

For confined water, our analyses of the ISF and MSD distinguish between water molecules in various pore regions. Figure \ref{fig1} shows $\ln \tau_\alpha$ and $1/r_{p}^2$ as a function of the shortest distance $d$ between a considered oxygen atom and any atom of the pore wall at the beginning of the time interval. Although different liquid-matrix interactions are absent, there is a strong slowdown of the $\alpha$ relaxation near neutral walls \cite{Vogel_13,Vogel_14}. Interestingly, the inverse amplitude of the cage rattling increases in a similar way when approaching such wall. In Fig.\ \ref{fig1}(c), we see that the data for most pore regions and temperatures collapse onto a master curve in a plot $\ln \tau_\alpha(1/r_{p}^2)$. For glass-forming bulk liquids, such collapse is predicted by elastic models, e.g., the shoving model, which proposes that the elastic distortion of the surroundings determines the activation energy of a flow event \cite{Dyre_06}. Notable deviations from the master curve are observed at low temperatures and, in particular, near the interface, specifically, for the three largest values of $1/r_{p}^2$ corresponding to water molecules at distances $d<0.3\,$nm. We suppose that elastic models do not adequately describe the hopping motion, which prevails in this pore region \cite{Vogel_13}, and strive for an extension of these models. 

\begin{figure}
 \centering
 \includegraphics*[width=5.5cm]{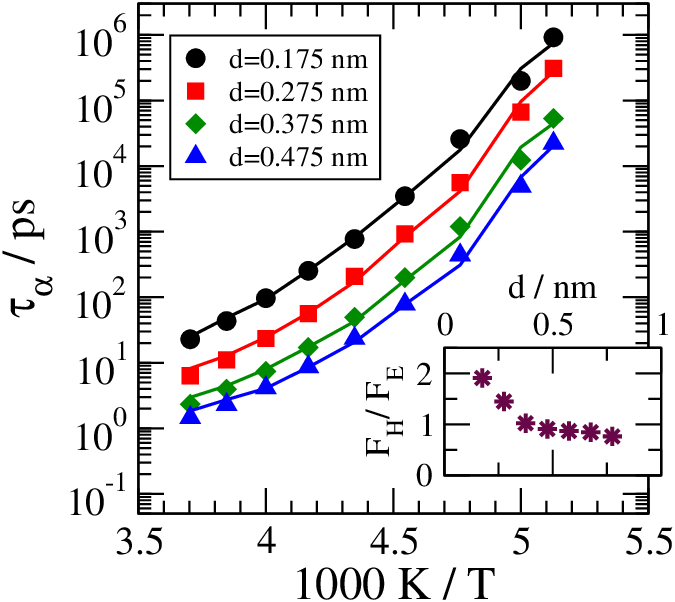}
\caption{Temperature dependence of $\tau_\alpha$ at various distances to the pore wall $d$. The solid lines are fits using Eq.\ \eqref{fitgleichung}. Inset: Resulting ratio of hopping barrier ($F_H$) and elastic barrier ($F_E$) at 210\,K in various pore regions.}
\label{fig2}
\end{figure}

In order to consider both elastic and hopping contributions, we proceed with an ECNLE approach to confined water. This theory proposes that, at sufficiently low temperatures, long-time and short-time dynamics are related according to  \cite{Mirigian_14-1} 
\begin{equation}
 \ln \left(\frac{\tau_\alpha}{\tau_0}\right) \approx \frac{F_H + F_E}{k_BT} \approx \sqrt{\frac{A}{r_p^2}}+ \frac{B\cdot \xi^2}{r_p^2}
\label{fitgleichung}
\end{equation}
Thus, the first and second terms on the right hand side of Eq. (\ref{fitgleichung}) are estimates for the hopping ($F_H$) and elastic ($F_E$) contributions to the free energy barrier $F$, respectively. Throughout this Letter, we use $\tau_0 = 0.18 \text{ ps}$, but we ensured that our conclusions are not affected when this parameter is varied in a reasonable range. The length scale $\xi$ describes the growth of a cooperative volume $V_c$ upon cooling, explicitly, $V_c(T)\propto \xi^2(T)$ within ECNLE theory \cite{Mirigian_13, Mirigian_14-1}. 

Here, we assume that $\xi(T)$ can be identified with or is, at least, proportional to the dynamic and static length scales determined for confined water in our previous studies \cite{Vogel_13, Vogel_14}. There, we used the dynamic profiles $\ln \tau_\alpha (d)$ to obtain a dynamic length $\xi_d$ up to which the dynamics of the liquid molecules are significantly affected by the fixed molecules. Furthermore, we studied persistent density correlations to extract a static length $\xi_s$ up to which the molecular positions in the confined liquid are substantially correlated with that in the confining matrix. We found that, widely independent of the size and geometry of the neutral confinement, $\xi_d\approx\xi_s$ grow essentially linearly with $1/T$ from $0.15\,$nm at 300\,K to $0.31\,$nm at 200\,K \cite{Vogel_14}. Exploiting this fact, we fix the length scale $\xi(T)$ at $\xi_d(T)\approx\xi_s(T)$ in all present analyses. Thus, there are only two free parameters, $A$ and $B$, to describe the temperature- and position-dependent dynamics of confined water.

Figure \ref{fig2} depicts the temperature dependence of $\tau_\alpha$ in various pore regions. In addition to a slowdown of the dynamics, a reduction of the fragility is observed when approaching the pore wall \cite{Vogel_13}. Fitting the data to Eq.\ \eqref{fitgleichung}, we obtain nice interpolations for all spatial regions. In particular, the present ECNLE approach captures both the slower dynamics and the reduced fragility of interfacial water. To characterize the relevance of the hopping and elastic contributions to the energy barrier in various confinement regions, the ratio $F_H/F_E$ is shown in the inset. We observe that, relative to the elastic contribution $F_E$, the hopping contribution $F_H$ is larger at smaller distances $d$ to the pore wall. Thus, the reduced fragility of interfacial water can be attributed to more jump-like dynamics near the wall, consistent with our previous observation that the prevailing motional mechanism differs in various confinement regions \cite{Vogel_13}.   

\begin{figure}
 \centering
 \includegraphics*[width=7.5cm]{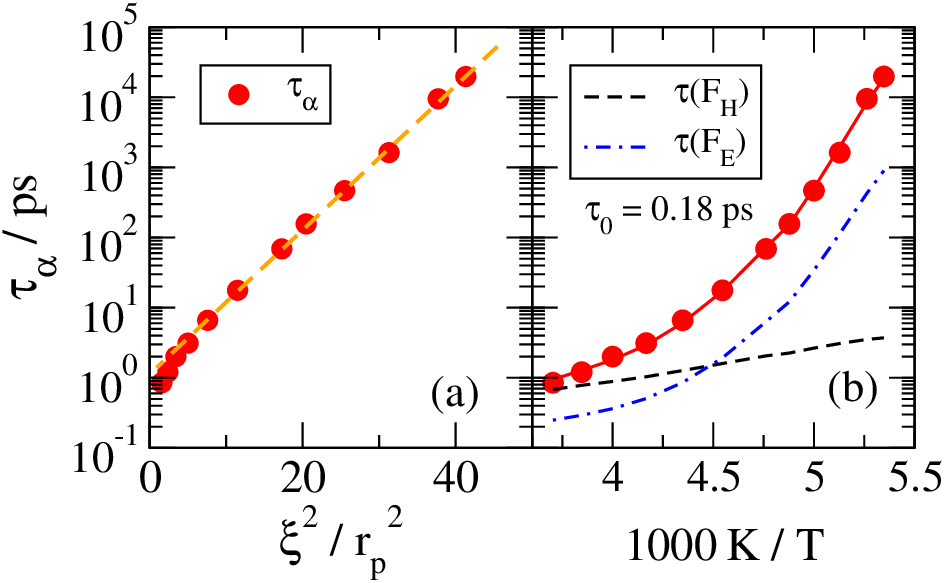}
\caption{Correlation times $\tau_\alpha$ of bulk water as a function of (a) $\xi^2(T)/r^2_p(T)$ and (b) $1/T$. In panel (a), the straight line is a guide to the eye. In panel (b), the solid line is a fit to Eq.\ \eqref{fitgleichung}; the dashed lines are results for $\tau_\alpha$ calculated considering only (black) $F_\text{H}$ or (blue) $F_\text{E}$ in this ECNLE approach.}
\label{fig3}
\end{figure}

Next, we switch to bulk water. First, we ascertain to which extent the glassy slowdown can be described by the present ECNLE approach. Figure \ref{fig3} shows the pronounced temperature dependence of $\tau_\alpha$ for bulk water. In Fig.\ \ref{fig3}(a), we investigate the relation between long-time and short-time dynamics. Evidently, $\ln \tau_\alpha$ is a linear function of $\xi^2(T)/r_p^2(T)$ at low temperatures, i.e., at large values of this ratio, while there is a different behavior at high temperatures. Motivated by these findings, we use Eq.\ \eqref{fitgleichung} to fit also the temperature evolution of $\tau_\alpha$ for bulk water. In Fig.\ \ref{fig3}(b), we see that the ECNLE approach describes the glassy slowdown. In all these analyses, $\xi(T)$ was again identified with $\xi_d(T) \approx \xi_s(T)$ from the studies of confined water \cite{Vogel_14}. Since the latter approach was limited to $T\geq 195\,$K, the length scales at lower temperatures were obtained from a mild extrapolation of the linear increase with $1/T$ at higher temperatures. Hence, our ECNLE approach involves merely two free parameters, $A$ and $B$. Figure \ref{fig3}(b) also displays values of $\tau_\alpha$ calculated from Eq.\ \eqref{fitgleichung} for the hypothetical situations that only the hopping contribution ($B=0$) or the elastic contribution ($A=0$) existed. A comparison of these values reveals that the energy barrier is dominated by $F_H$ and $F_E$ in the high-temperature and low-temperature ranges, respectively. Moreover, we see that the local hopping barrier yields an Arrhenius-like temperature dependence, while the non-local elastic barrier causes the high fragility at lower temperatures. It is important to note that fits using only the hopping contribution or the elastic contribution were not successful. Likewise, it was not possible to obtain a reasonable interpolation based on a temperature-independent length scale $\xi$. 

\begin{figure}
 \centering
 \includegraphics*[width=8.5cm]{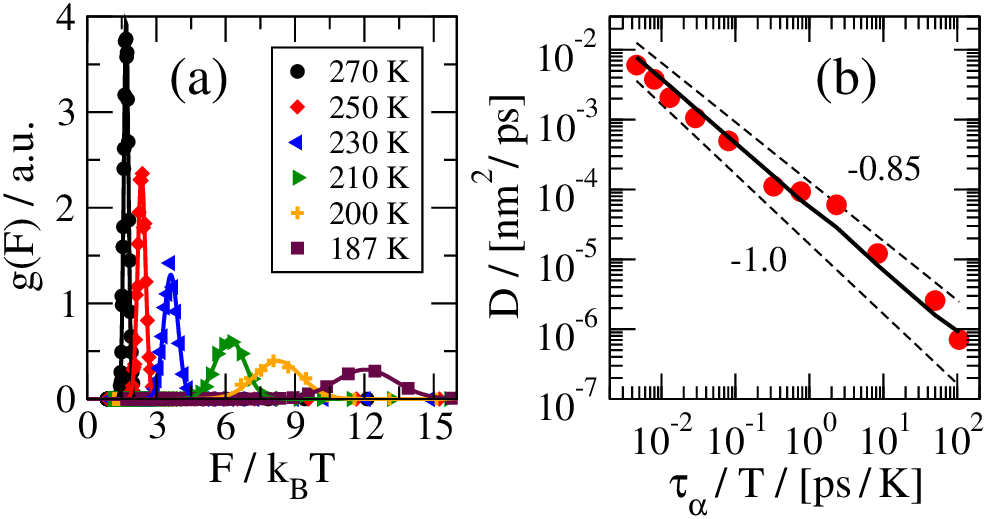}
\caption{(a) Distribution of reduced barriers $\frac{F}{k_BT}=\frac{F_B+F_E}{k_BT}$, as obtained from the squared displacements due to cage rattling, $r^2(t=1\,\mathrm{ ps})$, using Eq.\ \eqref{fitgleichung} and the parameters $A$ and $B$ from the fit in Fig.\ \ref{fig3}(b). Lines: Interpolations with Gaussian distributions. (b) $D$ as a function of $\tau_\alpha/T$. Solid line: Prediction for $D(\tau_\alpha/T)$ based on Eqs.\ \eqref{stokesein} and \eqref{gamma}. The constant of proportionality was adjusted to match the data. Dashed lines: Relations for $\gamma=0.85$ and $\gamma=1.0$, respectively. }
\label{fig4}
\end{figure}

Having demonstrated an intimate relation between $\tau_\alpha$ and $1/r_{p}^2$ for confined and bulk waters, we proceed to ascertain to what extent the present ECNLE approach allows us to rationalize the characteristic SE breakdown of glass-forming liquids \cite{Angell_00, Cavagna_09, Rossler_90, Cicerone_96, Fujara_94, Ediger_03, Starr_06, Starr_07, Sastry_13}. The SE relation links the self-diffusion coefficient $D$ to the shear viscosity $\eta$, explicitly, $D\eta/T = \text{const}$. While such behavior is found in regular liquids, fractional rather than linear SE relations were reported for supercooled liquids \cite{Fujara_94, Ediger_03, Sastry_13}, including supercooled water \cite{Starr_06, Starr_07}. Hereby, several studies exploited the empirical relation between $\eta$ and $\tau_\alpha$ and studied the fractional behavior  
\begin{equation} 
D \propto (\tau_\alpha/T)^{-\gamma}  \hspace{1.5 cm}(\gamma<1)
\label{stokesein} 
\end{equation}
This SE breakdown was often rationalized in terms of spatially heterogeneous dynamics in combination with the fact that different observables reflect different types of averages over the distributed motional states \cite{Ediger_00, Tarjus_95, Schweizer_04}. 

To analyze the SE breakdown, we exploit that, in an ECNLE framework, distributions of short-time squared displacements, $r^2(t=1\,\mathrm{ps})$, provide access to distributions of effective barriers $g(F)$. Specifically, using Eq.\ \eqref{fitgleichung} together with the parameters $A$ and $B$ from the above ECNLE analysis of $\tau_\alpha(T)$, the individual displacements due to cage rattling are mapped onto effective barriers $F=F_B+F_E$. In Fig.\ \ref{fig4}(a), we see that there are broad distributions of effective barriers, which amount to several $k_BT$ in the studied temperature range. Thus, there are effective barriers of significant height, as assumed for the derivation of Eq.\ \eqref{fitgleichung} \cite{Mirigian_14-1, Mirigian_14-2}. When decreasing the temperature, the reduced barrier heights $F/T$ grow more rapidly than expected from the temperature change. The same is true for the width of their distribution. Hence, the effective barriers $F$ grow and their distribution broadens upon cooling. Interpolations indicate a Gaussian shape of $g(F)$ with temperature-dependent values of mean $\overline{F}$ and variance $\sigma^2$. 

When dynamics are governed by Gaussian distributions $g(F)$, one expects exponents \cite{Schweizer_04}
\begin{equation}
\gamma(T)= \left(1-\frac{\sigma^2(T)}{2 \overline{F}(T)} \right)\left(1+\frac{\sigma^2(T)}{2\overline{F}(T)}\right)^{-1}
\label{gamma}
\end{equation}
To ascertain whether an SE breakdown can be traced back to an increasing dynamical heterogeneity associated with the broadening of the barrier distribution, we determine the self-diffusion coefficients of the water oxygens from the simulated long-time MSD, $\langle r^2(t\!\rightarrow\!\infty)\rangle=6Dt$. Figure \ref{fig4}(b) shows $D(\tau_\alpha/T)$. Despite relatively mild deviations in our case of water, the relation is not described by the SE exponent $\gamma=1$. To rationalize the SE breakdown, we employ Eq.\ \eqref{gamma} to calculate $\gamma(T)$. Using the resulting values in Eq.\ \eqref{stokesein}, we compute a prediction $D(T)$ from $\tau_\alpha(T)/T$, where the temperature-independent constant of proportionality is chosen to match the directly determined self-diffusion coefficients. In Fig.\ \ref{fig4}(b), it is evident that this prediction agrees with the simulation results for $D(\tau_\alpha/T)$. Thus, we conclude that the present ECNLE approach well describes the SE breakdown of the bulk water model. 

It should be noted that ECNLE theory obtains the displacement-dependent dynamic free energy from the static structure \cite{Mirigian_14-1, Mirigian_14-2}. Since neutral confinements do not alter the static structure \cite{Vogel_13, Vogel_14}, regular ECNLE approaches do not capture the slowdown of molecular dynamics near such interfaces. Our approach does not rely on an explicit calculation of the free energy from the static structure, but it uses the general relation between long-time and short-time dynamics predicted by the ECNLE theory. One may speculate that free-energy calculations for confined liquids should not use the static structure, but rather consider structural fluctuations due to cage rattling.

In summary, it was shown that the slowdowns upon confining or cooling water can be understood on common grounds within an ECNLE approach \cite{Mirigian_13}, which proposes a coupling of elastic distortions and hopping motion. In particular, we observed an intimate relation of short-time and long-time dynamics predicted by the theory. Furthermore, we found that interfaces or densification lead to an obstruction of cage rattling, which, in turn, results in a slowdown of structural rearrangements. Use of neutral confinements enabled a determination of a length scale $\xi$, which characterizes a cooperative volume of a structural relaxation event. This length scale roughly doubles in the studied temperature range \cite{Vogel_14}. Utilizing $\xi(T)$ in our ECNLE approach, it was possible to describe the temperature- and position-dependent dynamics of confined water as well as the glassy dynamics of bulk water, including the SE breakdown. We expect that our findings apply not only to water, but also to other confined and bulk glass-forming liquids. It remains a rewarding objective for future studies to what extent the present ECNLE approach applies to liquids dynamics in non-neutral confinement.

Funding of the Deutsche Forschungsgemeinschaft (DFG) through Grant No.\ VO 905/9-2 is gratefully acknowledged.

\end{document}